\documentclass[preprint,aps,pre]{revtex4}
\usepackage{epsfig}
\usepackage{latexsym}
\begin{document}

\title{Directed Spiral Site Percolation on the Square Lattice}

\author{S. B. Santra}

%\address{Department of Physics, Indian Institute of Technology,\\ 
%Guwahati-781039, Assam, India}
\affiliation{Department of Physics, Indian Institute of Technology,
Guwahati\\ Guwahati-781039, Assam, India}

\date \today

\begin{abstract}
A new site percolation model, directed spiral percolation (DSP), under
both directional and rotational (spiral) constraints is studied
numerically on the square lattice. The critical percolation threshold
$p_c\approx 0.655$ is found between the directed and spiral
percolation thresholds. Infinite percolation clusters are fractals of
dimension $d_f\approx 1.733$. The clusters generated are anisotropic.
Due to the rotational constraint, the cluster growth is deviated from
that expected due to the directional constraint. Connectivity lengths,
one along the elongation of the cluster and the other perpendicular to
it, diverge as $p\rightarrow p_c$ with different critical
exponents. The clusters are less anisotropic than the directed
percolation clusters. Different moments of the cluster size
distribution $P_s(p)$ show power law behavior with $|p-p_c|$ in the
critical regime with appropriate critical exponents. The values of the
critical exponents are estimated and found to be very different from
those obtained in other percolation models. The proposed DSP model
thus belongs to a new universality class. A scaling theory has been
developed for the cluster related quantities. The critical exponents
satisfy the scaling relations including the hyperscaling which is
violated in directed percolation. A reasonable data collapse is
observed in favour of the assumed scaling function form of
$P_s(p)$. The results obtained are in good agreement with other model
calculations.

\end{abstract}
\maketitle

\section{introduction}
Recently, there is strong interest in studying properties of
electro-rheological and magneto-rheological fluids \cite{gandhi},
magnetic semiconductors \cite{msemi}, and composite materials
\cite{cmat} because of their industrial applications. A new site
percolation model, directed spiral percolation (DSP), proposed here
could be applied in general to study their rheological, electrical and
magnetic properties. The DSP model is constructed imposing both
directional and rotational constraints on the ordinary percolation
(OP) model \cite{prc}. The directional constraint is in a fixed
direction in space and the empty sites in that direction are
accessible to occupation. Due to the rotational constraint the sites
in the forward direction or in a rotational direction, say clockwise,
are accessible to occupation. The direction of the rotational
constraint is not fixed in space and it depends on the direction from
which the present site is occupied. For charged particles, the
directional constraint in the model could arise from an electric field
along a particular direction in the plane of the lattice and the
rotational constraint may arise from a magnetic field applied
perpendicular to the plane. The model will also be applicable to the
physical situations corresponding to the presence of other kinds of
directional and rotational constraints.

The effect of two different external constraints, directional
constraint and rotational (spiral) constraint, on the ordinary
percolation model have been studied independently. The corresponding
models are known as directed percolation (DP) model \cite{zallen} and
spiral percolation (SP) model\cite{rsb}. The DP model has applications
in self-organized criticality \cite{soc}, reaction diffusion
systems\cite{rdif}, nonlinear random resistor networks \cite{rrn}
etc. The SP model has been applied in studying spiral forest fire
\cite{ff}, pinning of interfaces \cite{pin}, and diffusion under
rotational bias in disordered systems \cite{cdif}. It is observed that
in the presence of an external constraint the critical properties of
the system as well as the universality class of the OP model are
changed \cite{bunde}. The critical exponents associated with different
cluster related quantities in the DP and SP models are not only
different but also different from those in the case of the OP
model. The DP and SP models then belong to two different universality
classes than that of the OP model.

In this paper, the proposed DSP model is studied numerically on the
square lattice in two dimensions ($2D$). The external constraints in
the model determine the nearest empty sites available for
occupation. A single cluster growth Monte Carlo (MC) algorithm is
developed for this model under both the external constraints. Critical
percolation probability $p_c$, at which a spanning (or infinite)
cluster appears for the first time, is determined. The spanning
clusters are found fractals. The clusters generated in the OP and the
SP models are isotropic. In DP, the clusters are anisotropic and
elongated along the directional constraint. The clusters generated in
the case of the DSP model are anisotropic but they grow in a direction
different from the directional constraint due to the presence of the
rotational constraint. A new type of cluster is thus generated in this
model. The connectivity length exponents $\nu_\parallel$ and
$\nu_\perp$ are estimated. $\nu_\parallel$ is approximately equal to
the connectivity exponent $\nu$ of OP but both $\nu_\parallel$ and
$\nu_\perp$ are different from those obtained in DP. The clusters are
less anisotropic than DP clusters. Different moments of the cluster
size distribution $P_s(p)$ become singular as $p\rightarrow p_c$ with
their respective critical exponents. The values of the critical
exponents obtained here are different from those obtained in other
percolation models like OP, DP, and SP. The DSP model then belongs to
a new universality class. A scaling theory is developed. The critical
exponents satisfy the scaling relations within error bars including
the hyperscaling which is violated in DP. A reasonable data collapse
is observed in support of the assumed scaling function form of
$P_s(p)$. The results obtained in this model are in good agreement
with the results obtained in the study of magnetoresistance in a model
of $3$-constituents composite material \cite{bbs}.

\section{The DSP Model}
\begin{figure}
\bigskip
\centerline{\hfill \psfig{file=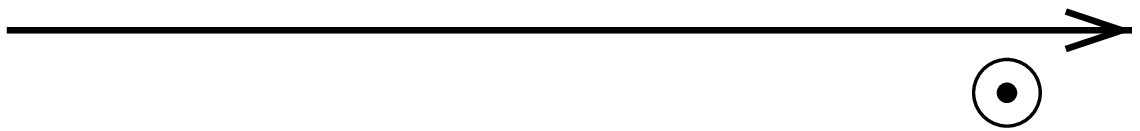,width=0.7\textwidth} \hfill}
\medskip
\centerline{\hfill \psfig{file=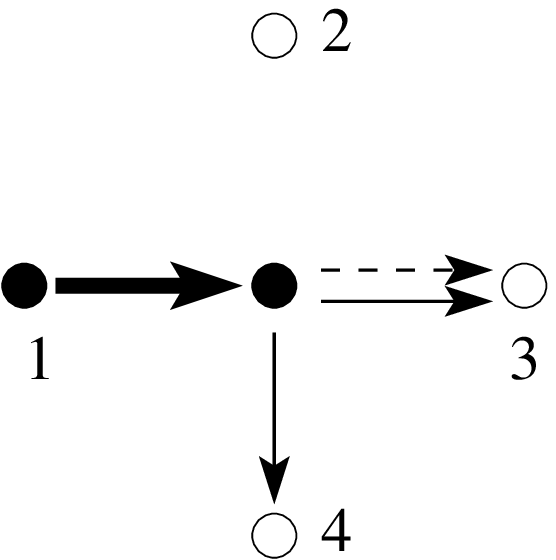,width=0.175\textwidth} \hfill\hfill
\psfig{file=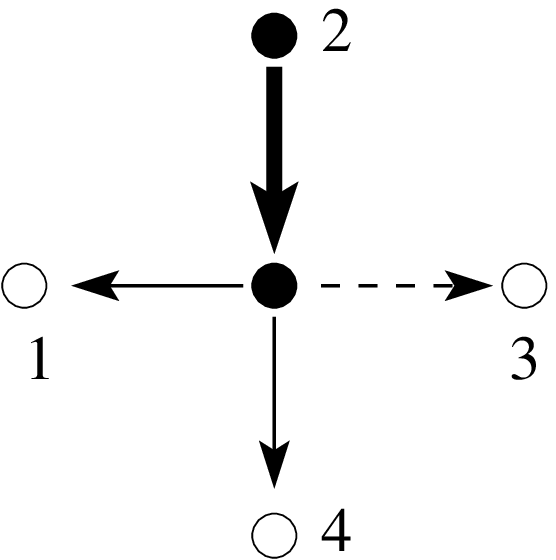,width=0.175\textwidth} \hfill\hfill
\psfig{file=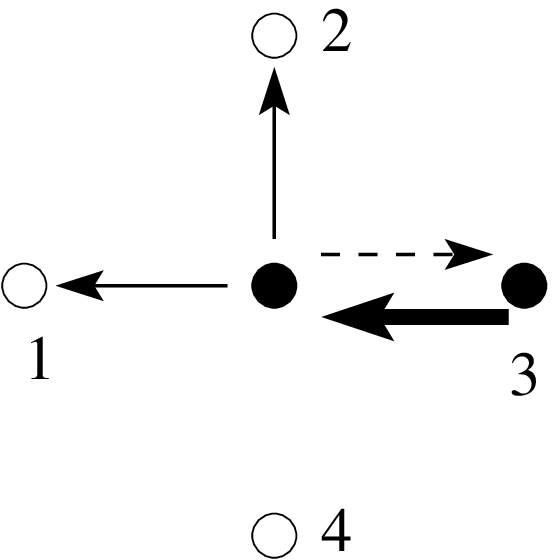,width=0.175\textwidth} \hfill\hfill
\psfig{file=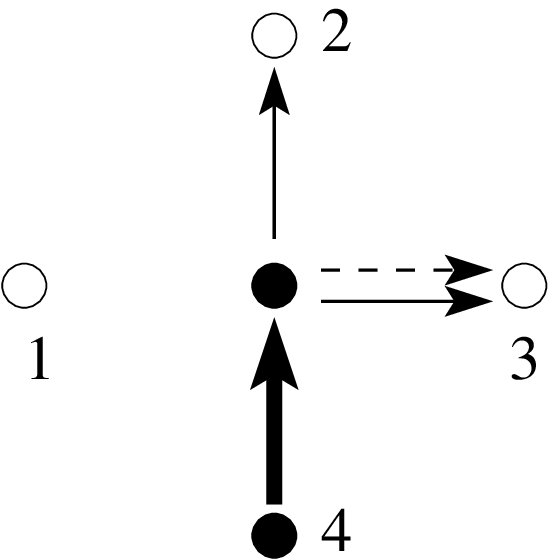,width=0.175\textwidth} \hfill }
\medskip
\centerline{\hfill \psfig{file=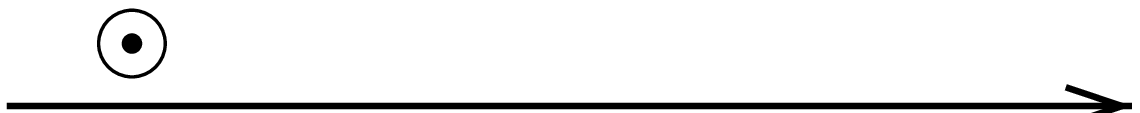,width=0.7\textwidth} \hfill}
\centerline{\hfill $(a)$ \hfill\hfill $(b)$ \hfill\hfill $(c)$ \hfill
\hfill $(d)$ \hfill}
\bigskip
\caption{\label{demo} Selection of empty nearest neighbors of an
occupied site for occupation. Black circles are the occupied sites and
open circles are the empty sites. Two thick long arrows from left to
right represent the directional constraint. The presence of clockwise
rotational constraint is shown by the encircled dots. The eligible
empty nearest neighbors of the central occupied site will be selected
here for occupation. The direction from which the central site is
occupied is indicated by a short thick arrow. The dotted arrow
indicates the sites allowed by the directional constraint and the thin
arrows indicate the sites allowed by the rotational constraint. In
$(a)$, the central site is occupied from the left and the sites $3$
and $4$ are accessible to occupation. In $(b)$, the central site is
approached from the top and sites $1$, $3$ and $4$ could be
occupied. In $(c)$, the central site is occupied from the right and
$1$, $2$, and $3$ are probable for occupation. Notice that the site
$3$, already an occupied site, could be reoccupied from a different
direction. In $(d)$, the central site is approached from the below and
sites $2$ and $3$ are the probable sites for occupation. }
\end{figure}

A square lattice of size $L\times L$ is considered.  A directional
constraint from left to right and a clockwise rotational constraint
are imposed on the system. Due to the directional constraint any empty
site on the right of an occupied site could be occupied in the next
Monte Carlo (MC) time step. Due to the rotational constraint the empty
sites in the forward direction or in the clockwise direction can be
occupied. To generate clusters under these two constraints a single
cluster growth algorithm is developed following the original algorithm
of Leath\cite{leath}. In this algorithm, the central site of the
lattice is occupied with unit probability. All four nearest neighbors
to the central site can be occupied with equal probability $p$ in the
first time step. As soon as a site is occupied, the direction from
which it is occupied is assigned to it. In the next MC time step, two
empty sites due to the rotational constraint and the site on the right
due to the directional constraint will be eligible for occupation for
each occupied site in the previous time step. This is illustrated in
Fig. \ref{demo}. The directional constraint is represented by two long
arrows from left to right. The presence of the rotational constraint
is shown by the encircled dots. The black circles represent the
occupied sites and the open circles represent the empty sites. The
empty nearest neighbors of the central occupied site will be selected
for occupation in this step. The direction from which the central site
is occupied is represented by a short thick arrow. The dotted arrow
indicates the eligible empty site for occupation due to directional
constraint and the thin arrows indicate the eligible empty sites for
occupation due to the rotational constraint. Since the directional
constraint is to the right, site $3$, an empty site on the right of
the occupied site, is always eligible for occupation. The sites
accessible to occupation due to the rotational constraint will be
identified now. The rotational constraint acts in the forward or in
the clockwise direction and depends on the direction of approach to
the present occupied site. In Fig. \ref{demo} $(a)$, the central site
is occupied from site $1$ on the left. Thus, site $3$ in the forward
direction and site $4$ in the clockwise direction are the eligible
sites for occupation due to the rotational constraint. In this
situation, sites $3$ and $4$ are then the only eligible sites for
occupation in the next time step due to both the constraints. In
$(b)$, the central site is occupied from site $2$ on the top and thus,
sites $4$ and $1$ are the eligible sites for occupation due to the
rotational constraint. The available sites for occupation are $1$, $3$
and $4$ due to both the constraints in this case. In $(c)$, the
eligible sites for occupation are $1$ and $2$ due to the rotational
constraint and site $3$ due to the directional constraint. Note that,
site $3$ is already an occupied site and could be reoccupied from a
different direction. A site is forbidden for occupation from the same
direction. On the square lattice, a site then could be occupied at
most $4$ times from $4$ possible directions. This is unlike the case
of the ordinary and directed percolations where a site is occupied
only once. In ($d$), sites $2$ and $3$ are eligible for occupation due
to both constraints. It can be seen that the direction of the
rotational constraint is not fixed in space and depends on the
previous time step whereas that of the directional constraint remains
fixed in space. In that sense, the directional constraint is a global
constraint and the rotational constraint is a local constraint in the
model. After selecting the eligible sites for occupation, they are
occupied with probability $p$. The coordinate of an occupied site in a
cluster is denoted by $(x$,$y)$. Periodic boundary conditions are
applied in both directions and the coordinates of the occupied sites
are adjusted accordingly whenever the boundary is crossed. At each
time step the span of the cluster in the $x$ and $y$ directions $L_x =
x_{max} - x_{min}$ and $L_y = y_{max} - y_{min}$ are determined. If
$L_x$ or $L_y\ge L$, the system size, then the cluster is considered
to be a spanning cluster. The critical percolation probability $p_c$
is defined as below which there is no spanning cluster and at $p=p_c$
a spanning cluster appears for the first time in the system.

A typical spanning or infinite cluster generated on a $128\times 128$
square lattice at $p=0.655$ is shown in Fig.\ref{cluster}. The black
dots represent the occupied sites. The cross near to the upper left
corner is the origin of the cluster (it was the central site of the
lattice). The thick arrows from left to right at the top and bottom
represent the directional constraint. The encircled points in the
upper right and lower left corners represent the existence of the
clockwise rotational constraint. Notice that all the dangling ends are
clockwisely turned as it is expected. The cluster is highly
rarefied. Holes of all possible sizes are there. The elongation of the
cluster is almost along the left upper to the right lower diagonal of
the lattice and not along the directional constraint applied in the
$x$ direction. The geometry of the infinite clusters in other
percolation models are the following. In OP, the clusters are
isotropic. In DP, the anisotropic clusters are elongated along the
applied field\cite{db}. In SP, the clusters are compact and
isotropic\cite{sb}. Not only the geometry but also the internal
structure, highly rarefied with curly dangling ends, of the cluster is
different from that of the infinite clusters in other percolation
models. A new type of anisotropic cluster is thus generated in the DSP
model.

\section{Scaling Theory}
The cluster related quantities and their singularities at $p=p_c$ are
defined here. Scaling relations among the critical exponents
describing the singularities of the cluster related quantities will
also be established. The cluster size distribution is defined as
\begin{equation}
\label{csd}
P_s(p)=\frac{N_s}{N_{tot}}
\end{equation}	
where $N_s$ is the number of $s$-sited clusters in a total of
$N_{tot}$ clusters generated. In the single cluster growth method, the
origin is occupied with unit probability. The scaling function form of
the cluster size distribution is then assumed to be
\begin{equation}
\label{scalef}
P_s(p)=s^{-\tau+1}{\sf f}[s^\sigma(p-p_c)]
\end{equation}
where $\tau$ and $\sigma$ are two exponents.

The order parameter of the percolation transition is the probability
that a site belongs to a spanning (or infinite) cluster and here it is
defined as $P_\infty=p-p\sum_s'P_s(p)$. The sum is over all the finite
clusters and it is indicated by a prime. The leading singularity of
$P_\infty$ will be governed by $\sum'_sP_s(p)$, the zeroth moment of
cluster size distribution. $P_\infty$ goes to zero as $p\rightarrow
p_c$ from the above with a critical exponent $\beta$. The exponent
$\beta$ is defined as
\begin{equation}
\label{expnpi}
P_\infty\sim (p-p_c)^\beta.
\end{equation}

An important cluster related quantity is the average cluster size
$\chi$. In the single cluster growth method it is the first moment of
the cluster size distribution $P_s(p)$ and defined as $\chi = \sum'_s
sP_s(p)$ where the sum is over the finite clusters. Two next higher
moments $\chi_1$ and $\chi_2$ are also defined as $\chi_1 = \sum'_s
s^2P_s(p)$ and $\chi_2 = \sum'_s s^3P_s(p)$. In the present context
these higher moments have no physical meaning but might be useful in
particular situations. The moments $\chi$, $\chi_1$, and $\chi_2$
diverge with their respective critical exponent $\gamma$, $\delta$,
and $\eta$ at $p=p_c$. The critical exponents $\gamma$, $\delta$, and
$\eta$ are defined as
\begin{equation}
\label{expnm}
\hfill \chi \sim |p-p_c|^{-\gamma},
\hspace{0.2cm} \chi_1 \sim |p-p_c|^{-\delta}, \hspace{0.2cm} \&
\hspace{0.2cm} \chi_2 \sim |p-p_c|^{-\eta}. \hfill
\end{equation}
Since the cluster related quantities are just different moments of the
cluster size distribution function $P_s(p)$ then the critical
exponents associated with them are not all independent. All the
critical exponents could be expressed in terms of the exponents $\tau$
and $\sigma$ needed to describe $P_s(p)$ (Eq.\ref{scalef}). It could be
shown that the $k$th moment of $P_s(p)$ become singular as
\begin{equation}
\label{scalee}
\Sigma'_ss^kP_s(p)\sim (p-p_c)^{-(k-\tau+2)/\sigma}.
\end{equation} 
Putting the value of $k$, the order of the moment, in Eq.\ref{scalee}
one could obtain the following scaling relations
\begin{equation}
\label{relexp1}
\hfill \beta=(\tau-2)/\sigma, \hspace{0.2cm} \gamma=(3-\tau)/\sigma,
\hspace{0.2cm} \delta=(4-\tau)/\sigma, \hspace{0.2cm} \&
\hspace{0.2cm} \eta=(5-\tau)/\sigma. \hfill
\end{equation}
Eliminating $\tau$ and $\sigma$, scaling relations between $\beta$,
$\gamma$, $\delta$, and $\eta$ could be obtained as
\begin{equation}
\label{relexp2}
\delta=\beta+2\gamma \hspace{0.5cm} \& \hspace{0.5cm}
\eta=2\delta-\gamma.
\end{equation}

Since the clusters generated here are anisotropic in nature, two
lengths, $\xi_\parallel$ and $\xi_\perp$, are needed to describe the
connectivity of the occupied sites.  $\xi_\parallel$ is the
connectivity length along the elongation of the cluster and
$\xi_\perp$ is the connectivity length along the perpendicular
direction to the elongation. To measure $\xi_\parallel$ and
$\xi_\perp$ the moment of inertia tensor ${\bf T}$, a $2\times 2$
matrix here, is calculated. For a $s$-sited cluster, the $xy$
component of the tensor is given by
\begin{equation}
\label{tcomp}
T_{xy}=\sum^s_{\ell=1}(x_{\ell}-x_0)(y_{\ell}-y_0)
\end{equation}
where $x_{\ell}$ and $y_{\ell}$ are the $x$ and $y$ coordinates of the
$\ell$th site and $(x_0$,$y_0)$ is the coordinate of the center of
mass of the cluster. The radii of gyration $R_\parallel(s)$ and
$R_\perp(s)$ with respect to two principal axes could be obtained as
$R^2_\parallel(s) = \lambda_1/s$ and $R^2_\perp(s) = \lambda_2/s$
where $\lambda_1$ is the largest eigenvalue and $\lambda_2$ is the
smallest eigenvalue of the $2\times 2$ moment of inertia matrix ${\bf
T}$. $R_\perp$ is about the axis passing through $(x_0$,$y_0)$ and
along the elongation of the cluster and $R_\parallel$ is about the
axis perpendicular to the elongation and passing through
$(x_0$,$y_0)$. The connectivity lengths now can be determined as
\begin{equation}
\label{concl}
\xi_\parallel^2=\frac{2\sum'_sR^2_\parallel(s) sP_s(p)}{\sum'_ssP_s(p)},
\hspace{0.5cm} \& \hspace {0.5cm}
\xi_\perp^2=\frac{2\sum'_sR^2_\perp(s) sP_s(p)}{\sum'_ssP_s(p)}.
\end{equation}
The correlation lengths $\xi_\parallel$ and $\xi_\perp$ diverge with
two different critical exponents $\nu_\parallel$ and $\nu_\perp$ as
$p\rightarrow p_c$. The critical exponents $\nu_\parallel$ and
$\nu_\perp$ are defined as
\begin{equation}
\label{expnc}
\xi_\parallel\sim |p-p_c|^{-\nu_\parallel} \hspace{0.5cm} \& \hspace {0.5cm}
\xi_\perp\sim |p-p_c|^{-\nu_\perp}.
\end{equation} 
The cluster mass is given by the number of sites $s$ in the cluster
and is expected to scale as $s \approx R_\parallel R_\perp^{(d_f-1)}$
at $p=p_c$, and it should go as $s \approx R_\parallel
R_\perp^{(d-1)}$ above $p_c$, where $d$ is the spatial dimension of
the lattice and $d_f$ is the fractal dimension of the infinite
clusters generated on the same lattice. The percolation probability
$P_\infty$ is the ratio of the number of sites on the infinite cluster
to the total number of sites, 
\begin{equation}
\label{pinfr}
P_\infty=\frac{R_\parallel R_\perp^{(d_f-1)}}{R_\parallel R_\perp^{(d-1)}}
\end{equation}
for $R_\parallel < \xi_\parallel$ and $R_\perp < \xi_\perp$. Assuming
$R_\parallel \sim \xi_\parallel$ and $R_\perp \sim \xi_\perp$, two
hyperscaling relations could be obtained as
\begin{equation}
\label{relexp3}
\nu_\perp(d_f-1) + \nu_\parallel=\frac{1}{\sigma} \hspace{0.5cm} \& \hspace
{0.5cm} (d-1)\nu_\perp - \beta = \nu_\perp(d_f-1).
\end{equation}
Eliminating $d_f$ from the above relations another scaling relation
could be obtained as
\begin{equation}
\label{relexp4}
(d-1)\nu_\perp+\nu_\parallel=\gamma+2\beta=2\delta-3\gamma=3\eta-4\delta.
\end{equation} 

In the following, the values of the critical exponents will be
determined and the scaling relations will be verified.

\section{Results and Discussions} 
\begin{figure}
\bigskip
\centerline{\hfill \psfig{file=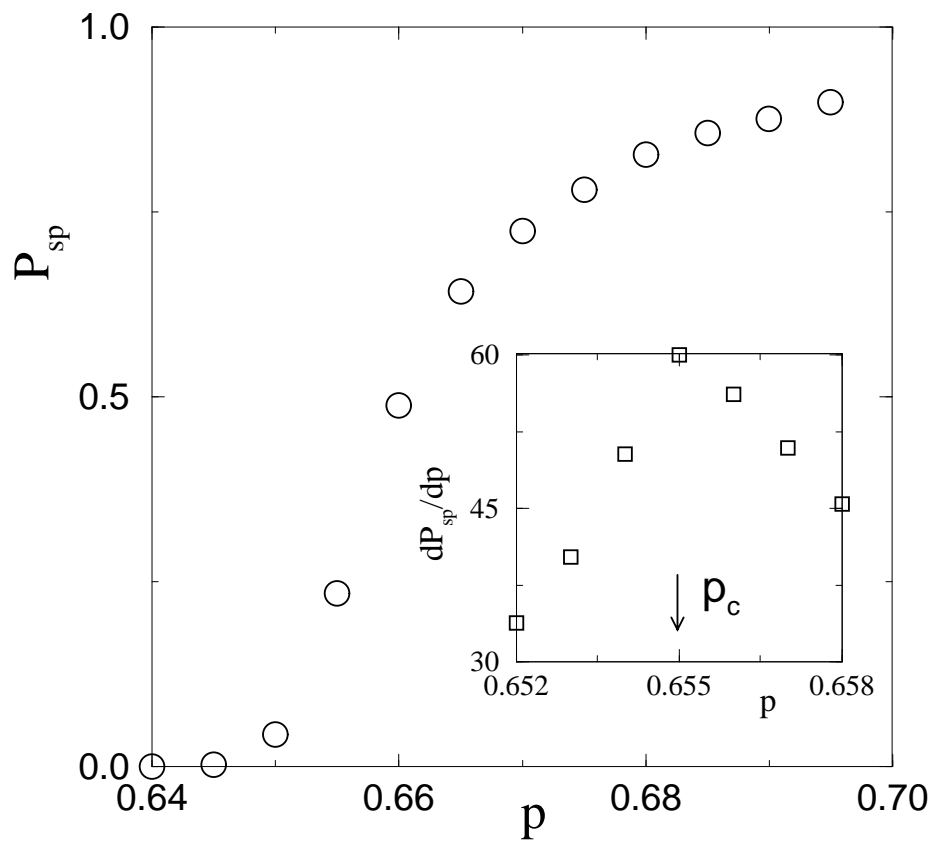,width=0.75\textwidth} \hfill}
\bigskip
\caption{\label{pspan} Plot of spanning probability $P_{sp}$ versus
$p$ (circles). In the inset the slope $dP_{sp}/dp$ is plotted against
$p$ (squares). The critical probability $p_c$ is determined from the
maximum slope and it is found $p_c=0.655\pm 0.001$ as indicated by an
arrow.}
\end{figure}

Simulation is performed on a square lattice of size $2^{10} \times
2^{10}$. The critical probability $p_c$ at which a spanning cluster
appears for the first time in the system is determined first. The
probability to have a spanning cluster is given by
\begin{equation}
\label{spp}
P_{sp}=\frac{n_{sp}}{N_{tot}}=1- \sum_s{'P}_s(p)
\end{equation}
where $n_{sp}$ is the number of spanning clusters out of total
$N_{tot}=10^4$ number of clusters generated. In Fig.\ref{pspan},
$P_{sp}$ is plotted against the probability of occupation $p$. Note
that, $P_{sp}$ is not going to zero sharply at a particular value of
$p$. This is due to the finite size of the lattice chosen here. The
critical probability $p_c$ is then determined from the maximum slope
of the curve $P_{sp}$ versus $p$. In the inset of Fig.\ref{pspan} the
slope $dP_{sp}/dp$ is plotted against $p$. It can be seen that the
threshold $p_c$ is at $0.655\pm 0.001$ corresponding to the maximum
slope. The derivative is calculated using the central difference
method for the data points collected in an interval of $0.001$. Note
that, the value of $p_c$ obtained here for the DSP model is slightly
above the value of $p_c$ of the directed percolation $p_c(DP)\approx
0.6445$ \cite{db} and below the spiral percolation threshold
$p_c(SP)\approx 0.712$ \cite{sb}. It is expected. Because, the
rotational constraint tries to make the clusters compact by
reoccupying the occupied sites of the directed clusters and, at the
same time, the directional constraint tries to elongate the compact
spiral clusters.

\begin{figure}
\bigskip
\centerline{\hfill \psfig{file=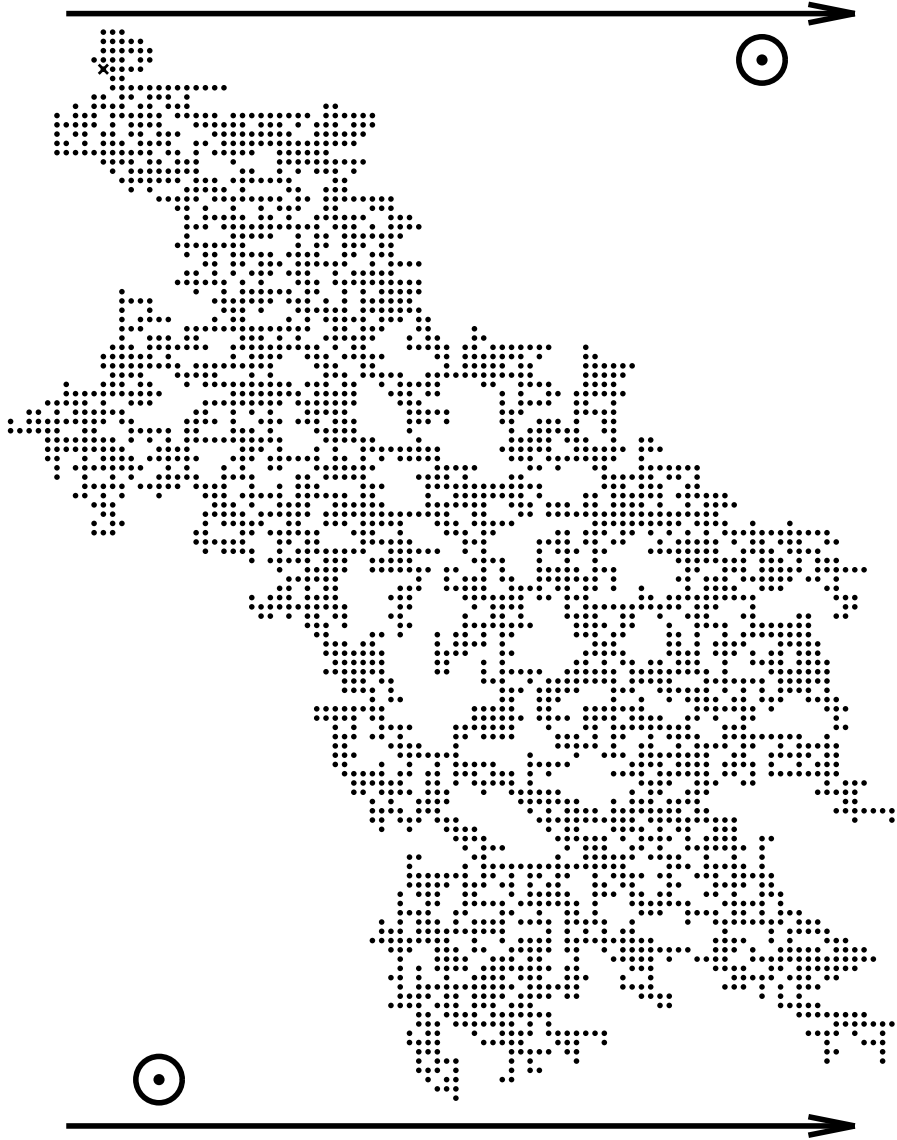,width=0.75\textwidth} \hfill}
\bigskip
\caption{\label{cluster} An infinite cluster on a $128\times 128$
square lattice at $p=0.655$ is shown. The black dots are the occupied
sites. The cross on the upper left corner is the origin of the
cluster. The thick arrows on the top and bottom from left to right
represent directional constraint. The presence of the clockwise
rotational constraint is shown by the encircled dots. The cluster is
highly rarefied and has holes of almost all possible sizes. The
elongation of the cluster is along the upper left to the lower right
diagonal and not along the directional constraint. The dangling ends
are clockwisely rotated. }
\end{figure}

\begin{figure}
\bigskip
\centerline{\hfill \psfig{file=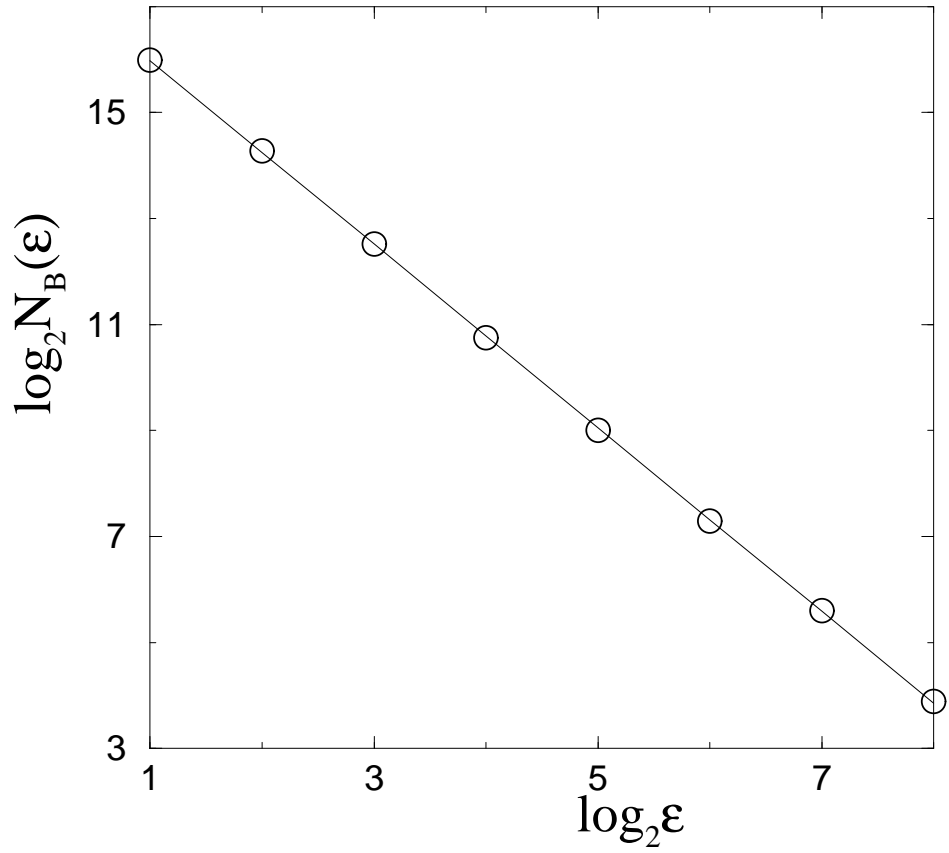,width=0.75\textwidth} \hfill}
\bigskip
\caption{\label{fracd} Number of boxes $N_B(\epsilon)$ is plotted
against the box size $\epsilon$. Data are averaged over $512$
samples. The fractal dimension is found $d_f=1.733\pm 0.005$. }
\end{figure}

The infinite cluster, shown in Fig. \ref{cluster}, generated on a
$128\times 128$ square lattice has holes of almost all possible
sizes. It seems that the infinite clusters are self-similar and
fractals. The fractal dimension $d_f$ of the infinite clusters
generated on the original $2^{10}\times 2^{10}$ lattice is determined
by the box counting method. The number of boxes $N_B(\epsilon)$ is
expected to grow with the box size $\epsilon$ as $N_B(\epsilon) \sim
\epsilon^{d_f}$ where $d_f$ is the fractal dimension. In
Fig. \ref{fracd}, $N_B(\epsilon)$ is plotted against the box size
$\epsilon$. The data are averaged over $512$ samples. A reasonably
good straight line is obtained in the $\log-\log$ scale. The fractal
dimension is found $d_f=1.733\pm 0.005$. The error is due to the least
square fitting of the data points taking into account the statistical
error of each point. The fractal dimension $d_f\approx 1.733$
$(\approx 12/7)$ obtained here is the smallest among the fractal
dimensions obtained in other percolation models. The values of $d_f$
in other percolation models are: $d_f(OP)= 91/48$ \cite{nn},
$d_f(DP)\approx 1.765$ \cite{hkv}, and $d_f(SP)\approx 1.957$
\cite{sb}. Also notice that the fractal dimension $d_f(DSP)$ obtained
here is little higher than the fractal dimension $1.64$ of ordinary
lattice animals\cite{lub}, large OP clusters below $p_c$. The infinite
cluster generated in the DSP model is then the most rarefied one among
the infinite clusters obtained in all four models.

\begin{figure}
\bigskip
\centerline{\hfill \psfig{file=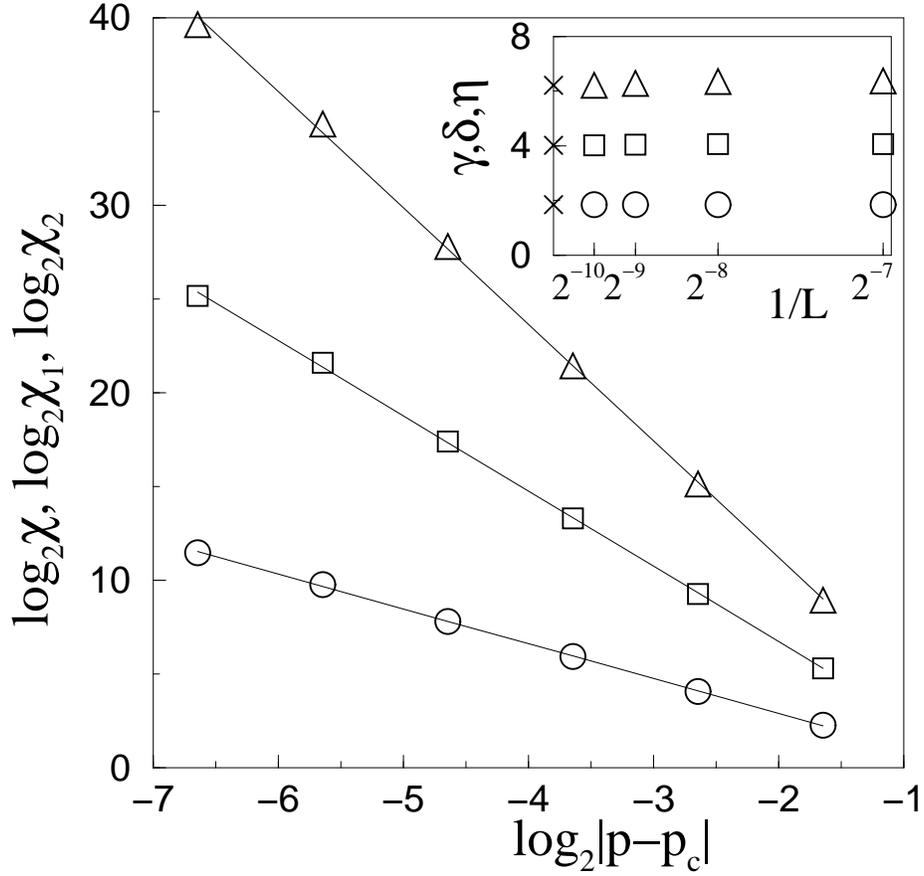,width=0.75\textwidth} \hfill}
\bigskip
\caption{\label{chi123} Plot of the first, second and third moments
$\chi$, $\chi_1$, and $\chi_2$ of the cluster size distribution versus
$|p-p_c|$. Different symbols are: circles for $\chi$, squares for
$\chi_1$, and triangles for $\chi_2$.  The corresponding critical
exponents are found as $\gamma=1.85\pm 0.01$, $\delta= 4.01\pm 0.04$,
and $\eta= 6.21\pm 0.08$. In the inset, the values of the exponents
$\gamma (\bigcirc)$ , $\delta (\Box)$ and $\eta (\triangle)$ are plotted
against the system size $1/L$. Extrapolating to the infinite system size
($1/L=0$), the values of the exponents obtained are $\gamma \approx
1.85$, $\delta \approx 4.01$, and $\eta \approx 6.21$.}
\end{figure}

Next, the values of the critical exponents $\gamma$, $\delta$, and
$\eta$ are estimated. The average cluster size $\chi$ and two other
higher moments $\chi_1$ and $\chi_2$ are measured generating $10^4$
finite clusters below $p_c$ for six different $p$ values. In
Fig. \ref{chi123}, $\chi$, $\chi_1$, and $\chi_2$ are plotted against
$|p-p_c|$. The circles represent $\chi$, the squares represent
$\chi_1$ and the triangles represent $\chi_2$. The values of the
exponents obtained are $\gamma = 1.85 \pm 0.01$, $\delta = 4.01 \pm
0.04$, and $\eta = 6.21 \pm 0.08$. The errors quoted here are the
standard least square fit error taking into account the statistical
error of each single data point. The values of the exponents $\gamma$,
$\delta$, and $\eta$ are also determined by the same Monte Carlo
technique on the square lattices of three different smaller sizes
$2^9\times 2^9$, $2^8\times 2^8$, and $2^7\times 2^7$ to check the
finite system size effects on the data. The values of the exponents
$\gamma$ (circles), $\delta$ (squares), and $\eta$ (triangles) for
different system sizes are plotted against the system size $1/L$ in
the inset of Fig. \ref{chi123}. It is then extrapolated upto
$L\rightarrow \infty$, the infinite system size. The extrapolated
values of the exponents are marked by crosses and the numerical values
obtained are $\gamma \approx 1.85$, $\delta \approx 4.01$, and $\eta
\approx 6.21$, the same as that of the system size $2^{10}\times
2^{10}$. These values of exponents are now used to verify the scaling
relations obtained in section III. One of the scaling relations $\eta
= 2\delta - \gamma$ obtained in Eq.\ref{relexp2} is checked now. The
value of $2\delta - \gamma = 6.17$ is very close to the value of the
exponent $\eta=6.21$. The scaling relation $\eta = 2\delta - \gamma$
is then satisfied within error bars. Using the scaling relation
$\delta = \beta + 2\gamma$ (Eq.\ref{relexp2}), the value of the
exponent $\beta$ is obtained as $\beta=0.31 \pm 0.06$. The error has
propagated from the errors of $\gamma$ and $\delta$. Independent
estimation of the exponent $\beta$ becomes difficult because of the
presence of curvature in the plot of $P_\infty$ versus $(p-p_c)$ in
the log-log scale for $p>p_c$. This may be due to dominating
corrections to scaling to the leading singularity of $P_\infty$. The
values of the exponents $\tau$ and $\sigma$ can also be estimated
using the scaling relations in Eq. \ref{relexp1}. Three different
values of $\tau$ could be obtained as $\tau_1 = (3\delta -
4\gamma)/(\delta - \gamma) = 2.14\pm 0.18$, $\tau_2 = (4\eta -
5\delta)/(\eta - \delta) = 2.18\pm 0.28$, and $\tau_3 = (3\eta -
5\gamma)/(\eta - \gamma) = 2.15\pm 0.11$. The estimate of $\tau$ can
be taken as the average of $\tau_1$, $\tau_2$ and $\tau_3$ and it is
given by $\tau= 2.16 \pm 0.20$. Similarly $\sigma= 0.459 \pm 0.015$ is
determined from $\sigma_1 = 1/(\delta - \gamma) = 0.463 \pm 0.011$,
$\sigma_2 = 1/(\eta - \delta) = 0.455 \pm 025$, and $\sigma_3 =
2/(\eta - \gamma) = 0.459\pm 0.009$. The errors quoted here are the
propagation errors. A comparison between the values of the exponents
obtained in different percolation models is made in Table
\ref{tab1}. It can be seen that the values of the exponents obtained
in this model are very different from those obtained in other
percolation models like OP, DP, and SP. The magnitude of $\beta$ is
the largest and the magnitude of $\gamma$ is the smallest among the
four models. The DSP model then belongs to a new universality
class. The values of the exponents obtained here could be approximated
to the nearest rational fractions as $\beta \approx 1/3$, $\gamma
\approx 11/6$, $\delta \approx 24/6$, $\eta \approx 37/6$, $\tau
\approx 28/13$ and $\sigma \approx 6/13$.  Surprisingly they satisfy
all the scaling relations in Eq.  \ref{relexp1} and Eq. \ref{relexp2}
exactly.

\begin{figure}
\bigskip
\centerline{\hfill \psfig{file=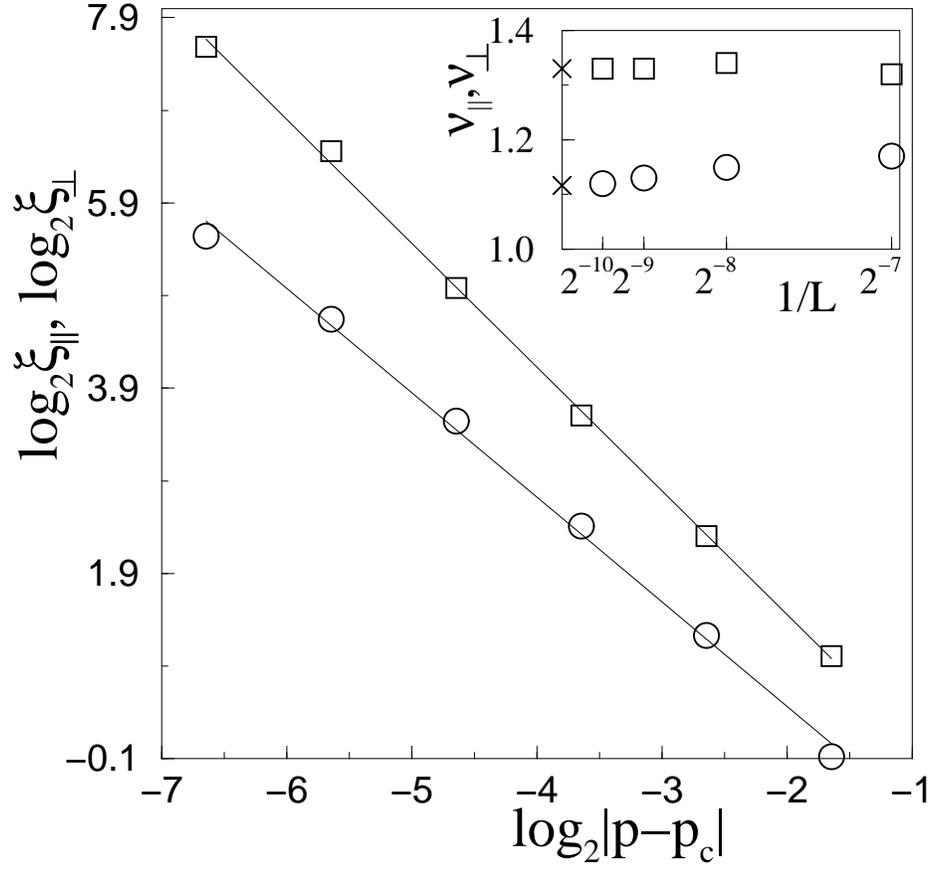,width=0.75\textwidth} \hfill}
\bigskip
\caption{\label{corrl} The connectivity lengths, $\xi_\parallel$ and
$\xi_\perp$, are plotted against $|p-p_c|$. The circles represent
$\xi_\perp$ and the squares represent $\xi_\parallel$. The critical
exponents are found as $\nu_\parallel= 1.33\pm 0.01$ and $\nu_\perp =
1.12 \pm 0.03$. In the inset, the values of $\nu_\parallel (\Box)$ and
$\nu_\perp (\bigcirc)$ are plotted against the system size $1/L$. The
values of the exponents are extrapolated upto $L\rightarrow
\infty$. The extrapolated values of the exponents are marked by
crosses and they are $\nu_\parallel \approx 1.33$ and $\nu_\perp
\approx 1.12$. }
\end{figure}

The connectivity lengths, $\xi_\parallel$ and $\xi_\perp$, for the
system size $2^{10}\times 2^{10}$ are plotted against $|p-p_c|$ in
Fig.\ref{corrl}. The squares represent $\xi_\parallel$ and the circles
represent $\xi_\perp$. The corresponding exponents $\nu_\parallel$ and
$\nu_\perp$ are obtained as $\nu_\parallel = 1.33\pm 0.01$ and
$\nu_\perp = 1.12\pm 0.03$. The errors quoted here are the least
square fit errors. In the inset of Fig. \ref{corrl}, $\nu_\parallel$
(squares) and $\nu_\perp$ (circles) are plotted against different
system sizes $1/L$.  The exponent values are extrapolated upto
$L\rightarrow \infty$ and the extrapolated values are $\nu_\parallel
\approx 1.33$ and $\nu_\perp \approx 1.12$, the same as obtained for
the system size $2^{10}\times 2^{10}$. There are few things to
notice. First, the value of $\nu_\parallel$ is almost equal to the
connectivity exponent of ordinary percolation, $\nu(OP) = 4/3$
\cite{nn}. Recently, a study of magnetoresistance of a three-component
composites consisting of cylindrical insulator and perfect conductors
in a metallic host film is made by Barabash et al \cite{bbs}. Since the
Hall effect will generate an electric field with a component
perpendicular to the plane of the film with an in-plane applied
current, their system appears inherently three dimensional
($3D$). However, the electric field perpendicular to the film plane
vanishes because of the presence of columnar perfect conductor. Thus,
their $3D$ problem reduces to that of calculating the effective
conductivity of a $2D$ composites of perfect insulator and perfect
conductor\cite{bbs,cmat}. The results obtained in the model of
composite material by Barabash et al \cite{bbs} then could be compared
with the results of the present DSP model in $2$ dimensions. It is
found by Barabash et al \cite{bbs} that the correlation length
exponent is $4/3$ independent of anisotropy. The correlation length
exponent quoted there is equivalent to the connectivity exponent
$\nu_\parallel$ of the DSP model considered here. The results of the
DSP model is thus in good agreement with the results obtained in the
model calculation of magnetoresistance of composite materials. Second,
the values of $\nu_\parallel$ and $\nu_\perp$ are different from those
obtained in the DP model, $\nu_\parallel(DP) = 1.733\pm0.001$ and
$\nu_\perp(DP) = 1.0972\pm 0.0006$ \cite{ebg} (see Table \ref{tab1}
also). Third, the ratio of the connectivity lengths goes as
$\xi_\parallel/\xi_\perp \sim |p-p_c|^{-\Delta\nu}$ where $\Delta\nu =
\nu_\parallel - \nu_\perp$. In DSP, $\Delta\nu$ is approximately
$0.21$ whereas in DP, it is approximately $0.64$. This means that
clusters in the DSP model are less anisotropic than the clusters in
the DP model at any $p$. This is due to the presence of the rotational
constraint which tries to make the clusters isotropic. Barabash et al
\cite{bbs} assumed that $\Delta\nu = 0$. The effective aspect ratio of
their problem might vary as $|p-p_c|^{-\Delta\nu}$ with $\Delta\nu =
0.21$ as obtained here in the DSP model. Fourth, $(d-1)\nu_\perp
+\nu_\parallel$ is $\approx 2.45$ for $d=2$ and $2\delta - 3\gamma
\approx 2.47$. The hyperscaling relation $2\delta - 3\gamma =
(d-1)\nu_\perp +\nu_\parallel$ in Eq. \ref{relexp4} is then satisfied
within error bars. Two other hyperscaling relations $\nu_\perp(d_f-1)
+ \nu_\parallel=1/\sigma$ and $(d-1)\nu_\perp - \beta =
\nu_\perp(d_f-1)$ given in Eq. \ref{relexp3} are also satisfied within
error bars. In directed percolation hyperscaling is violated
\cite{hp}. This is the first anisotropic percolation model where
hyperscaling is satisfied. The values of the connectivity exponents
$\nu_\parallel$ and $\nu_\perp$ could also be approximated to their
nearest rational fractions as $\nu_\parallel \approx 4/3$ and
$\nu_\perp \approx 7/6$. They satisfy the hyperscaling relations in
Eq. \ref{relexp3} and \ref{relexp4} exactly taking $d_f=12/7$.

In the above study, it is found that the fractal dimension $d_f$ and
the values of the critical exponents ($\tau$, $\sigma$, $\beta$,
$\gamma$, $\nu_\parallel$, $\nu_\perp$, etc.) of DSP model are
different from other percolation models. As a consequence, the DSP
model belongs to a new universality class. This is due to the fact
that a completely new type of percolation cluster is generated in the
DSP model. There are three important features of the DSP clusters. The
clusters are rarefied, have spiraling danging ends, and are
anisotropic. As the clusters grow, more and more vacancies are
generated into the cluster. This is because, the clockwise rotational
constraint tries to occupy sites away from the directional constraint
whereas the directional constraint tries to occupy sites along itself
in the $x$-direction. The effects of these two counter acting
constraints are the following: the spiral clusters (clusters with
spiraling dangling ends) become rarefied, become wider as it is away
from the origin (see Fig.\ref{cluster}), and elongated along a
clockwisely rotated direction from the original globally fixed
directional constraint. The clusters are neither directed percolation
clusters nor spiral percolation clusters. The combined directed and
spiral constraints produces highly rarefied anisotropic spiral
clusters. Individually, the effect of these three features, anisotrpoy
\cite{db}, spiraling \cite{sb}, and volume fraction \cite{kertez}, on
the ordinary percolation clusters have been studied. Each of them
corresponds to different critical behaviour and consequently belongs
to new universality class. Here in DSP model, a new critical behaviour
is obtained at the percolation threshold because of the presence of
all three features in the same cluster.

\begin{figure}
\bigskip
\centerline{\hfill \psfig{file=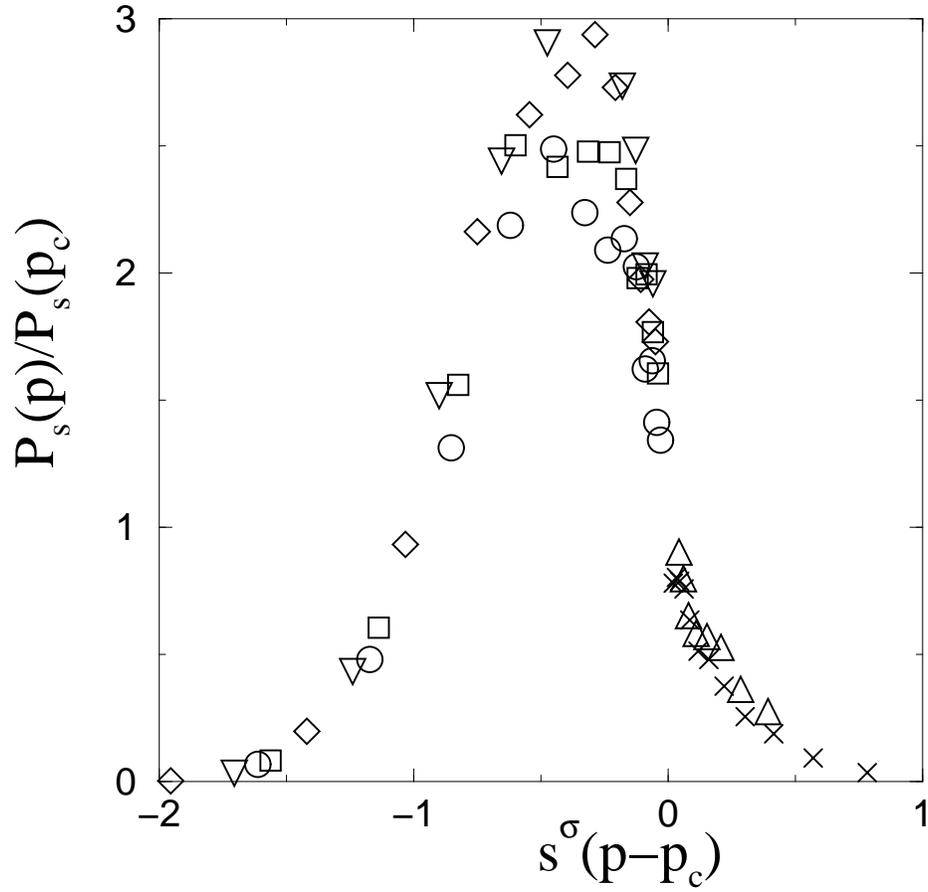,width=0.75\textwidth} \hfill}
\bigskip
\caption{\label{datac} Plot of the scaled cluster size distribution
$P_s(p)/P_s(p_c)$ versus the scaled variable $s^\sigma(p-p_c)$ for
different values of $p$ with $\sigma=0.459$. The cluster size $s$
changes from $64$ to $8192$. The data plotted correspond to $p-p_c=
0.12 (\times)$, $0.11 (\triangle)$, $-0.03 (\bigcirc)$, $-0.04
(\Box)$, $-0.05 (\Diamond)$, and $-0.06 (\bigtriangledown)$.  A
reasonable data collapse is observed. }
\end{figure}

Finally, the scaling function form assumed for the cluster size
distribution $P_s(p)=s^{-\tau+1}{\sf f}[s^\sigma(p-p_c)]$ is
verified. The scaled cluster size distribution $P_s(p)/P_s(p_c)$ is
now plotted against the scaled variable $s^\sigma (p-p_c)$ in
Fig. \ref{datac}. It is assumed that $\sf f[0]$ is a constant. The
value of $\sigma$ is taken as $0.459$. Distribution of $10^4$ clusters
over the bins of width $2^{i-1} - (2^i-1)$ with $i=1$ to $L^2$
$(=2^{20})$ is considered for each $p$ value. In Fig. \ref{datac}, the
cluster size $s$ changes from $64$ to $8192$ and $(p-p_c)$ changes
form $0.12$ to $-0.06$. Data for different values of $s$ and $(p-p_c)$
collapse reasonably onto a single curve. This means that the scaling
function form assumed for the cluster size distribution is appropriate
for this model. Also notice that the maximum value of
$P_s(p)/P_s(p_c)$ for DSP model ($\approx 3$) is different from that
of OP model ($\approx 4.5$) \cite{prc}.

\section{Conclusion}
The new directed spiral site percolation, DSP model, under both
directional and rotational constraints belongs to a new universality
class. The combined directed and spiral constraints produces a new
type of percolation clusters which are highly rarefied, anisotropic,
and spiral in nature. The critical properties of the cluster related
quantities in this model are very different from the other percolation
models like OP, DP, and SP. The critical exponent $\gamma$ of the
average cluster size $\chi$ is the smallest and the exponent $\beta$
of $P_\infty$ is the largest among the four models. The fractal
dimension $d_f\approx 12/7$ is the smallest among the four percolation
models and thus the infinite clusters are highly rarefied here. Since
the clusters generated in this model are anisotropic, two connectivity
lengths, $\xi_\parallel$ and $\xi_\perp$, are defined to describe the
scaling behaviour of the clusters connectivity as $p \rightarrow
p_c$. The exponents $\nu_\parallel$ and $\nu_\perp$, associated with
$\xi_\parallel$ and $\xi_\perp$, are estimated. The finite size effect
on the exponent values is also checked making simulations on different
system sizes and the values of the exponents are extrapolated to the
infinite network. It is found that $\nu_\parallel$ is approximately
equal to the connectivity exponent $\nu=4/3$ of the OP model which is
also in good agreement with the correlation length exponent $\nu$
obtained by Barabash et al \cite{bbs} in the study of
magnetoresistance of a model composite material. Both $\nu_\parallel$
and $\nu_\perp$ are different from those obtained in the DP model. The
order of anisotropy $\xi_\parallel / \xi_\perp$ is higher in the DP
clusters than that of the clusters in the DSP model at any $p$. The
critical exponents of the DSP model satisfy the scaling relations
within error bars including the hyperscaling which is violated in the
DP model. This is the first anisotropic percolation model where the
hyperscaling is satisfied. The assumed scaling function form $P_s(p) =
s^{-\tau+1} {\sf f} [s^\sigma(p-p_c)]$ of the cluster size
distribution $P_s(p)$ is verified through data collapse. The values of
all the critical exponents are suggested in terms of rational
fractions and it is observed that those rational fractions satisfy the
scaling relations exactly. The proposed rational fractions for the
values of the critical exponents might be verified through an exact
solution of the model. A difficulty in solving the model exactly is in
the reoccupation of the same sites from different directions. The
model will be applicable to the physical situations where both the
directional and rotational constraints are present. For example,
measurement of magnetoresistance in composite materials, magnetic
semiconductors, and super-ionic conductors could be studied using this
model if it is extended to $3$ dimensions. The model will also be
applicable in studying the rheological properties of
electro-rheological and magneto-rheological fluids in presence of
crossed electric and magnetic fields.

\section{acknowledgment}
The author thanks Indrani Bose for helpful discussions and critical
comments on the manuscript.

\newpage
\begin{table}
\begin{tabular}{p{2.5cm}p{2cm}p{2cm}p{2cm}p{2cm}p{2.5cm}p{2cm}}
\hline Percolation  & $\beta$ & $\gamma$ & $\tau$ & $\sigma$ &
$\nu$ & $d_f$\\ 
Models &&&&&&\\
\hline 

OP\cite{nn} & $5/36$ & $43/18$ & $187/91$ & $36/91$ & $4/3$ & $91/48$\\

DP\cite{hkv,ebg} & $0.277$ & $2.2772$ & $2.108$ & $0.3915$ &
$\nu_\perp=1.0972$ & $1.765$\\ & $\pm 0.002$ &$\pm 0.0003$ & $\pm
0.001$ & $\pm 0.0004$ & $\pm 0.0006$ & \\ &&&&& $\nu_\parallel=1.733$
&\\ &&&&& $\pm 0.001$ &\\

SP\cite{sb} & $0.048$ & $2.19$ & $2.022$ & $0.447$ & $1.116$ &
$1.957$\\ & $\pm 0.011$ & $\pm 0.07$ & $\pm 0.004$ & $\pm 0.014$ &
$\pm 0.003$ & $\pm 0.009$\\

DSP & $0.31$ & $1.85$ & $2.16$ & $0.459$ & $\nu_\perp=1.12$ & $1.733$
 \\ & $\pm 0.01$ & $\pm 0.01$ & $\pm 0.20$ & $\pm 0.004$ & $\pm 0.03$
 & $\pm 0.005$ \\ & $(1/3)$ & $(11/6)$ & $(28/13)$ & $(6/13)$ &
 $(7/6)$ & $(12/7)$ \\ &&&&& $\nu_\parallel=1.33$ &\\ &&&&& $\pm 0.01$
 &\\ &&&&& $(4/3)$ &\\
\hline
\end{tabular}
\bigskip
\caption{\label{tab1} Comparison of the values of the critical
exponents $\beta$, $\gamma$, $\tau$, $\sigma$, $\nu$, and $d_f$ in the
case of ordinary (OP), directed (DP), spiral (SP), and directed spiral
(DSP) percolation on the square lattice. The values of the critical
exponents of the DSP model are different from the other models. The
values within parenthesis are the nearest rational fractions of the
values of the critical exponents. These rational fractions satisfy the
scaling relations exactly. The DSP model belongs to a new universality
class. }
\end{table}
\newpage

\end{document}